\documentclass[conference]{IEEEtran}
\IEEEoverridecommandlockouts
\usepackage{cite}
\usepackage{amsmath,amssymb,amsfonts}
\usepackage{graphicx}
\usepackage{textcomp}
\usepackage{xcolor}
\def\BibTeX{{\rm B\kern-.05em{\sc i\kern-.025em b}\kern-.08em
    T\kern-.1667em\lower.7ex\hbox{E}\kern-.125emX}}

\usepackage{subfig}
\usepackage{graphicx}
\usepackage{graphics}
\usepackage{epstopdf}
\usepackage{amsmath}
\usepackage{multirow}
\usepackage{bm}
\usepackage{enumerate}
\usepackage{mathdots}
\usepackage{todonotes}
\usepackage{algorithmicx}
\usepackage{algpseudocode}
\usepackage{algorithm}
\usepackage{verbatim}
\usepackage{balance}
\usepackage{cite}
\usepackage{url}
\usepackage{acronym}

\newacro{ML}{{Maximum Likelihood}}
\newacro{CFO}{{carrier frequency offset}}
\newacro{TO}{timing offset}
\newacro{LLF}{{Log-Likelihood Function}}
\newacro{mmWave}{millimeter wave}
\newacro{FIM}{Fisher Information Matrix}
\newacro{CRLB}{Cram\'{e}r-Rao Lower Bound}
\newacro{SW-OMP}{Simultaneous Weighted - Orthogonal Matching Pursuit}
\newacro{AoA}{angles-of-arrival}
\newacro{AoD}{angles-of-departure}
\newacro{NMSE}{Normalized Mean Squared Error}
\newacro{CNMSE}{Cumulative Normalized Mean Squared Error}
\newacro{CSI}{channel state information}
\newacro{NR}{New Radio}
\newacro{PN}{phase noise}
\newacro{MAP}{Maximum A Posteriori}
\newacro{PSD}{power spectral density}

\usepackage{color}
\definecolor{purple(x11)}{rgb}{0.63, 0.36, 0.94}
\definecolor{cadmiumgreen}{rgb}{0.0, 0.42, 0.24}

\newcommand{\vect}{\mathop{\mathrm{vec}}}

\newcommand{\Real}{\mathop{\mathrm{Re}}}
\newcommand{\Imag}{\mathop{\mathrm{Im}}}

\newcommand{\diag}{\mathop{\mathrm{diag}}}

\newcommand{\SNR}{\mathop{\mathrm{SNR}}}

\usepackage{pifont}

\newcommand{\Ttr}{N_\mathrm{tr}}
\newcommand{\Lc}{L_\mathrm{c}}

\newcommand{\Ts}{T_{\mathrm{s}}}
\newcommand{\Nr}{N_{\mathrm{r}}}
\newcommand{\Nt}{N_{\mathrm{t}}}

\newcommand{\Ns}{N_{\mathrm{s}}}

\newcommand{\Lr}{L_{\mathrm{r}}}
\newcommand{\Lt}{L_{\mathrm{t}}}

\newcommand{\Gr}{G_{\mathrm{r}}}
\newcommand{\Gt}{G_{\mathrm{t}}}

\newcommand{\ar}{{\mathbf{a}}_{\mathrm{R}}}
\newcommand{\at}{{\mathbf{a}}_{\mathrm{T}}}

\newcommand{\bsfFrf}{{\bsfF}_{\text{RF}}}

\newcommand{\bsfWrf}{{\bsfW}_\text{RF}}

\newcommand{\be}{\begin{eqnarray}}
\newcommand{\ee}{\end{eqnarray}}

\def\bg{{\mathbf{g}}}

\def\bp{{\mathbf{p}}}

\def\br{{\mathbf{r}}}
\def\bs{{\mathbf{s}}}

\def\bv{{\mathbf{v}}}

\def\bx{{\mathbf{x}}}
\def\by{{\mathbf{y}}}

\def\b0{{\mathbf{0}}}

\def\bA{{\mathbf{A}}}
\def\bB{{\mathbf{B}}}
\def\bC{{\mathbf{C}}}

\def\bE{{\mathbf{E}}}
\def\bF{{\mathbf{F}}}

\def\bH{{\mathbf{H}}}
\def\bI{{\mathbf{I}}}

\def\bP{{\mathbf{P}}}

\def\bX{{\mathbf{X}}}
\def\bY{{\mathbf{Y}}}
\def\bZ{{\mathbf{Z}}}

\def\bsfA{\bm{\mathsf{A}}}

\def\bsfC{\bm{\mathsf{C}}}
\def\bsfD{\bm{\mathsf{D}}}

\def\bsfF{\bm{\mathsf{F}}}
\def\bsfG{\bm{\mathsf{G}}}
\def\bsfH{\bm{\mathsf{H}}}
\def\bsfI{\bm{\mathsf{I}}}

\def\bsfP{\bm{\mathsf{P}}}

\def\bsfS{\bm{\mathsf{S}}}

\def\bsfW{\bm{\mathsf{W}}}

\def\sfg{{\mathsf{g}}}

\def\sfj{{\mathsf{j}}}

\def\sfs{{\mathsf{s}}}

\def\sf0{{\mathsf{0}}}

\def\bsfff{{\bm{\mathsf{f}}}}
\def\bsfg{{\bm{\mathsf{g}}}}

\def\bsfq{{\bm{\mathsf{q}}}}

\def\bsfv{{\bm{\mathsf{v}}}}
\def\bsfw{{\bm{\mathsf{w}}}}

\def\bsfz{{\bm{\mathsf{z}}}}
\def\bsf0{{\bm{\mathsf{0}}}}

\begin{document}
\title{Joint Synchronization, Phase Noise and Compressive Channel Estimation in Hybrid Frequency-Selective mmWave MIMO Systems}
\author{Javier Rodr\'{i}guez-Fern\'{a}ndez$^{\dag}$ and Nuria Gonz\'{a}lez-Prelcic$^{\dag\ddag}$ \\
$^\dag$ The University of Texas at Austin, Email: $\{$javi.rf,ngprelcic$\}$@utexas.edu and $^\ddag$ University of Vigo}
\maketitle

\begin{abstract}
The large beamforming gain used to operate at \ac{mmWave}  frequencies requires obtaining channel information to configure hybrid antenna arrays. Previously proposed wideband channel estimation strategies, however, assume perfect time-frequency synchronization and neglect phase noise, making these approaches impractical. Consequently, achieving time-frequency synchronization between transmitter and receiver and correcting for \ac{PN} as the channel is estimated, is  the greatest challenge yet to be solved in order to configure hybrid precoders and combiners in practical settings. In this paper, building upon our prior work, we  find the \ac{MAP} solution to the joint problem of \ac{TO}, \ac{CFO}, \ac{PN} and compressive channel estimation for broadband \ac{mmWave} MIMO systems with hybrid architectures. Simulation results show that, using significantly less training symbols than in the beam training protocol in the 5G New Radio communications standard, joint synchronization and channel estimation at the low SNR regime can be achieved, and near-optimum data rates can be attained.
 \end{abstract}

%

\section{Introduction}


Hybrid MIMO architectures at \ac{mmWave} comprise of reasonably large antenna arrays to obtain the beamforming gain needed to compensate for the small antenna aperture resulting from moving towards high frequency bands. As analyzed in prior work \cite{RiaRusPreAlkHea:Hybrid-MIMO-Architectures:16}, they provide a reasonable trade-off between achievable performance and power consumption, and in a broad variety of cases enable attaining as high data rates as conventional all-digital MIMO architectures. MmWave MIMO links, however, need to be configured at low SNR regime, which sets the need to jointly synchronize and acquire \ac{CSI} under phase noise.


Most of prior work on channel estimation at \ac{mmWave}, however, assumes perfect synchronization at the receiver side \cite{RodGonVenHea:TWC:18}, \cite{VenAlkGon:Channel-Estimation-for-Hybrid:17}, \cite{JSTSP_2018}, \cite{ChEstHybridPrecmmWave}, \cite{Marzi2016:ChanEstTracking}. Prior work on channel estimation under synchronization impairments for \ac{mmWave} MIMO focuses on a narrowband channel model \cite{MyersH17a,RodGon:CFO_Channel_est:TWC:18}. An analog-only architecture is assumed in \cite{MyersH17a}, while \cite{RodGon:CFO_Channel_est:TWC:18} considers a hybrid MIMO system but does not address the problems of frame synchronization and phase noise compensation, which are crucial to establish synchronization. The problem of joint \ac{CFO}, \ac{PN}, and channel estimation has been previously studied in \cite{TSP_PN}, and \cite{Channel_PN_HCRLB}, although the proposed strategy operates under a SISO setting, with a single transmitted OFDM training symbol, and at very high SNR regime. When considering a frequency-selective scenario and a hybrid mmWave MIMO architecture,  to the best of our knowledge, only our previous work \cite{Joint_Sync_Chan_Est_Asilomar_2018} deals with the problem of designing a joint time-frequency synchronization and channel estimation strategy. The derived algorithms operate, however,  under the assumption that the \ac{PN} can be neglected.


In this paper, we consider the joint problem of \ac{TO}, \ac{CFO}, \ac{PN}, and \ac{mmWave} MIMO channel estimation, deriving a strategy that leverages the hybrid training precoder and combiner design in \cite{Joint_Sync_Chan_Est_Asilomar_2018}. Similarly to \cite{RodGon:CFO_Channel_est:TWC:18}, the channel sparsity level is assumed unknown, while the noise variance has to be previously obtained, since it is necessary to  estimate the \ac{PN} efficiently. TO, CFO, PN and low dimension beamformed channel are first estimated. Then, the \ac{SW-OMP} algorithm \cite{RodGonVenHea:TWC:18} is used to reconstruct the \ac{mmWave} MIMO channel from the beamformed channel. To the best of our knowledge, this is the first work that considers all the different synchronization impairments at \ac{mmWave} for channel estimation using hybrid architectures, proposing a design that is evaluated under the 5G \ac{NR} channel model. Numerical results show the effectiveness of the proposed estimation framework in terms of \ac{NMSE}, probability of detection, and achievable spectral efficiency. 

\section{System model}

\begin{figure*}[ht!]
\centering
\includegraphics[width=0.76\textwidth]{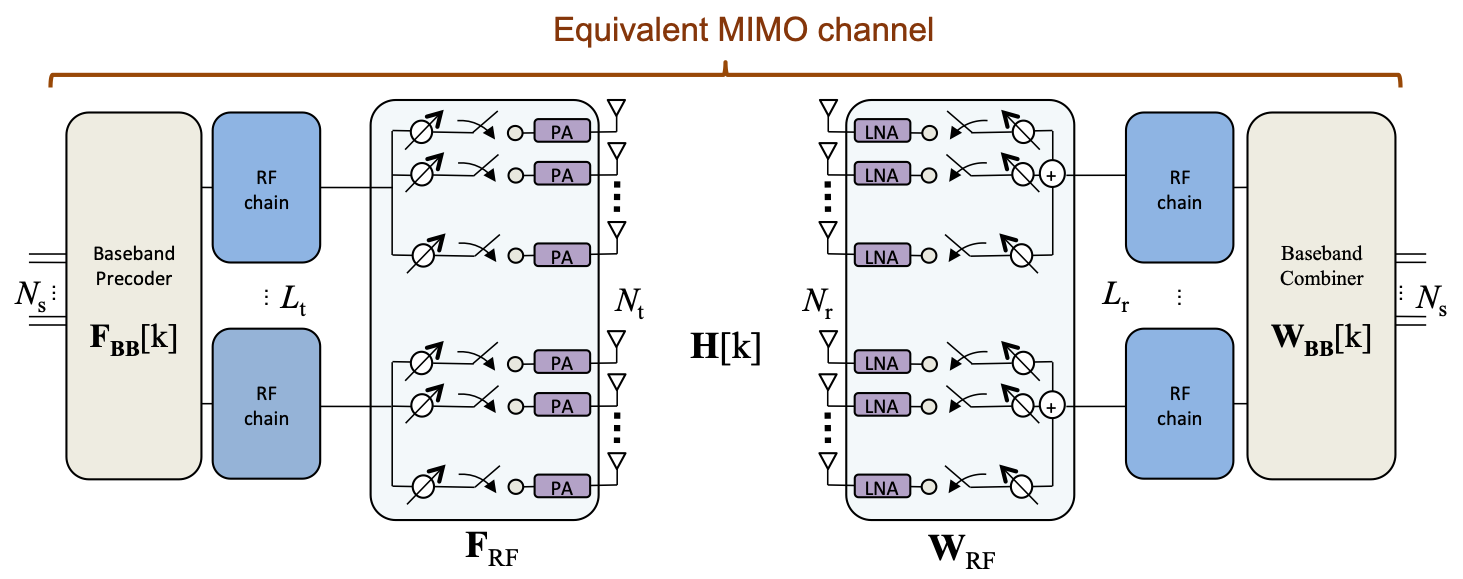} 
\caption{Illustration of the structure of a partially connected hybrid MIMO architecture, which includes analog and digital precoders and combiners.}     
\label{fig:hybrid_architecture}        
\end{figure*}

We consider a single-user \ac{mmWave} MIMO-OFDM communications link in which a transmitter equipped with $\Nt$ antennas sends a $\Ns \times 1$ vector $\bs^{(m)}[n]$ of data streams to a receiver having $\Nr$ antennas, with $m$ denoting the number of transmitted vectors.
Both transmitter and receiver are assumed to use a partially-connected hybrid MIMO architecture as shown in Fig. \ref{fig:hybrid_architecture}, with $\Lr$ and $\Lt$ RF chains. A frequency-selective hybrid precoder is used, with $\bsfF^{(m)}[k]= \bsfF_\text{RF}^{(m)}\bsfF_\text{BB}^{(m)}[k] \in {\mathbb{C}}^{\Nt\times\Ns}$, where $\bsfF_\text{RF}^{(m)} \in \mathbb{C}^{\Nt \times \Lt}$ is the analog precoder and $\bsfF_\text{BB}^{(m)}[k] \in \mathbb{C}^{\Lt \times \Ns}$ is the digital one at subcarrier $k$, $0 \leq k \leq K-1$. The RF precoder and combiner are implemented using a partially-connected network of phase-shifters, as described in \cite{Rial_switchOrShifter:Access2016}.

The frequency-selective MIMO channel between the transmitter and the receiver is modeled as a set of $\Nr \times \Nt$ matrices denoted as $\bH[d]$, $d = 0,\ldots,D-1$, with $D$ the delay tap length of the channel. Each of the matrices $\bH[d]$ is assumed to be a sum of the contributions of $C$ spatial clusters, each contributing with $R_c$ rays,  $c = 1,\ldots,C$. We use $\rho$ to denote the pathloss, $\alpha_{c,r} \in \mathbb{C}$ is the complex gain of the $r$-th ray within the $c$-th cluster, $\tau_{c,r} \in \mathbb{R}_{+}$ is the time delay of the $r$-th ray within the $c$-th cluster, $\phi_{c,r}, \theta_{c,r} \in [0,2\pi)$ are the \ac{AoA} and \ac{AoD}, and $\ar(\phi_{c,r}) \in {\mathbb{C}}^{\Nr\times1}$ and $\at(\theta_{c,r}) \in {\mathbb{C}}^{\Nt\times1}$ denote the receive and transmit array steering vectors. Let $p_\text{RC}(\tau)$ denote the equivalent pulse shape plus analog filtering effects, and $\Ts$ denotes the sampling interval. Using this notation, the channel matrix at delay tap $d$ is given by \cite{schniter_sparseway:2014}
\begin{equation}
\begin{split}
\bH[d] &= \sqrt{\frac{\Nr\Nt}{\rho \sum_{c=1}^{C}{R_c}}}\sum_{c = 1}^{C}\sum_{r=1}^{R_c}\alpha_{c,r} p_\text{RC}(d\Ts - \tau_{c,r}) \times \\ &\times \ar(\phi_{c,r})\at^*(\theta_{c,r}). 
\end{split}
\label{eqn:channel_model}
\end{equation}
The channel matrix can also be compactly represented in the frequency domain as \cite{RodGonVenHea:TWC:18}
\begin{equation}
\begin{split}
	\bsfH[k] &= \bsfA_\text{R} \bsfG[k] \bsfA_\text{T}^*,
\end{split}
\label{eq:channel_compact_fd}
\end{equation}
where $\bsfG[k] \in \mathbb{C}^{\sum_{c=1}^{C}{R_c} \times \sum_{c=1}^{C}{R_c}}$ is a diagonal matrix containing the path gains and the equivalent pulse-shaping effect, and $\bsfA_\text{T} \in \mathbb{C}^{\Nt \times \sum_{c=1}^{C}R_c}$, $\bsfA_\text{R} \in \mathbb{C}^{\Nr \times \sum_{c=1}^{C}R_c}$ are the array response matrices evaluated on the \ac{AoD} and \ac{AoA}, respectively.
Finally, we can approximate the matrix $\bsfH[k]$ in \eqref{eq:channel_compact_fd} using the extended virtual channel model \cite{mmWavetutorial} as
\begin{equation}
	\bsfH[k] \approx \tilde{\bsfA}_\text{R} \bsfG^\text{v}[k] \tilde{\bsfA}_\text{T}^*,
	 \label{equation:channel_extended_fd}
\end{equation}
where $\bsfG^\text{v}[k] \in \mathbb{C}^{\Gr \times \Gt}$ is a sparse matrix containing the path gains of the quantized spatial frequencies in the non-zero elements, and the dictionary matrices $\tilde{\bsfA} _\text{T} \in \mathbb{C}^{\Nt \times \Gt}$, $\tilde{\bsfA}_\text{R} \in \mathbb{C}^{\Nr \times \Gr}$ contain the transmit and receive array response vectors evaluated on spatial grids of sizes $\Gt$ and $\Gr$.

The receiver applies a hybrid combiner $\bsfW^{(m)}[k] = \bsfW_\text{RF}^{(m)} \bsfW_\text{BB}^{(m)}[k] \in {\mathbb{C}}^{\Nr\times\Lr}$, with $\bsfW_\text{RF}^{(m)} \in \mathbb{C}^{\Nr \times \Lr}$ the analog combiner, and $\bsfW_\text{BB}^{(m)}[k] \in \mathbb{C}^{\Lr \times \Ns}$ the baseband combiner at the $k$-th subcarrier. Let us assume that both the hybrid precoder and combiner are equally designed for all subcarriers, i.e., $\bsfW^{(m)}[k] = \bsfW^{(m)}$ and $\bsfF^{(m)}[k] = \bsfF^{(m)}$, $k = 0,\ldots,K-1$. The motivation is that, as shown in  \cite{RodGonVenHea:TWC:18}, the use of fequency-flat training precoders and combiners has been shown to be optimal in terms of preserving the Fisher Information. In this paper, we adopt the Zadoff-Chu-sequence based precoder and combiner design method in \cite{Joint_Sync_Chan_Est_Asilomar_2018}, which has been shown to provide an excellent trade-off between synchronization performance and compressive estimation capabilities. Now, let $n_0 \in \mathbb{K}_+$, $\Delta f^{(m)} \in \mathbb{R}$, $\theta[n] \in \mathbb{R}$ denote the unknown \ac{TO}, \ac{CFO} normalized to the sampling rate, and random phase shift experienced by the $n$-th received baseband sample. Then, the received signal at discrete time instant $n$ can be written as
\vspace*{-2mm}
\begin{equation}
\begin{split}
\br^{(m)}[n] &=  e^{\sfj \theta^{(m)}[n]}\left(\sum_{d=0}^{D-1}\bsfW^{(m)*}\bH[d]\bsfF^{(m)}\bs^{(m)}[n-d-n_0]\right) \\ &\times e^{\sfj 2\pi \Delta f^{(m)} n} + \bv^{(m)}[n],
\end{split} 
\label{equation:rx_signal}
\end{equation}
for $n = 0,\ldots,N+D+n_0-1$, with $N$ being the length of the time-domain sequence $\bs^{(m)}[n-d]$, and $\bv[n] \sim {\cal CN}\left(\bm 0, \sigma^2\bsfW^*\bsfW\right)$ is the post-combining received noise. In this paper, similarly to prior work \cite{RodGonVenHea:TWC:18}, we exploit the $\Lt$ available degrees of freedom coming from the transmit RF chains. Let $\bsfq^{(m)} \in \mathbb{C}^{\Lt \times 1}$ be a frequency-flat complex spatial modulation vector built from energy-normalized QPSK constellation symbols. Therefore, we will assume that $\bs^{(m)}[n]$ is of the form
\begin{equation}
	\bs^{(m)}[n] = \bsfq^{(m)} s^{(m)}[n],
\end{equation}
with $s^{(m)}[n] \in \mathbb{C}$ being the set of $\Ttr$ OFDM symbols that the $m$-th training frame comprises of. This signal can be expressed as
\begin{equation}
\begin{split}
	s^{(m)}[n] &= \frac{1}{K}\sum_{k=0}^{K-1}\sum_{t = 0}^{\Ttr-1}\sfs_t^{(m)}[k] e^{\sfj \frac{2\pi k (n - \Lc - t(K+\Lc))}{K}}, \\ & \quad n = 0,\ldots,(\Ttr-1)(K+\Lc)-1.
\end{split}
\label{equation:OFDM_t_signal}
\end{equation}



Let us consider the Cholesky decomposition of $\bsfC_\text{w}^{(m)}$ as $\bsfC_\text{w}^{(m)} = \bsfD_\text{w}^{(m)*}\bsfD_\text{w}^{(m)}$, with $\bsfD_\text{w}^{(m)} \in \mathbb{C}^{\Lr \times \Lr}$ an upper triangular matrix. Now, let us define a vector $\bg^{(m)}[d] \in \mathbb{C}^{\Lr \times 1}$, $\bg^{(m)}[d] = \bsfD_\text{w}^{(m)-*}\bsfW_\text{RF}^{(m)*} \bH[d] \bsfF_\text{RF}^{(m)}\bsfq^{(m)}$, containing the complex equivalent beamformed channel samples for a given training step $1 \leq m \leq M$. Accordingly, for the $m$-th transmitted frame, the received signal in \eqref{equation:rx_signal} can be expressed as
\begin{equation}
\begin{split}
	\br^{(m)}[n] &= e^{\sfj (2\pi \Delta f^{(m)} n + \theta^{(m)}[n])} \sum_{d=0}^{D-1} \underbrace{\bg^{(m)}[d] s^{(m)}[n-d-n_0]}_{\bx^{(m)}[n,d,n_0]} + \\ &+ \bv^{(m)}[n],
\end{split}
	\label{equation:rx_frame}
\end{equation}
with $\bv^{(m)}[n] \sim {\cal CN}(\bm 0, \sigma^2 \bI_{\Lr})$ being the post-whitened spatially white received noise vector, and $\bg^{(m)}[d] = [\alpha_1[d] e^{j\beta_1[d]},\ldots,\alpha_{\Lr}[d] e^{j\beta_{\Lr}[d]}]^T$ is the complex equivalent beamformed channel for the $m$-th training step and $d$-th delay tap. Let $\bm \theta^{(m)} \in \mathbb{R}^{\Ttr K \times 1}$ denote the phase noise samples experienced by the time-domain received symbols corresponding to the training subcarriers. The \ac{PN} model for IEEE 802.11ad is given in \cite{Draft_IEEE_802.11_PN}, whose \ac{PSD} is given in \cite{Blind_PN_ICC_2015_mmWave} as
\begin{equation}
P(f) = G_\theta \left[\frac{1 + (f/f_\text{z})^2}{1 + (f/f_\text{p})^2}\right],
\label{equation:PSD_PN}
\end{equation}
in which $G_\theta = -85$ dBc/Hz, $f_\text{z} = 100$ MHz, and $f_\text{p} = 1$ MHz \cite{Blind_PN_ICC_2015_mmWave}. Using the inverse Fourier transform of the \ac{PSD} in \eqref{equation:PSD_PN}, we can obtain the autocorrelation of the phase noise as
\begin{equation}
\begin{split}
R_{\theta^{(m)}\theta^{(m)}}&(\tau^{(m)}) = \mathbb{E}\{ \theta(t) \theta(t+\tau^{(m)}) \} \\
&= G_\theta \left[ \frac{f_\text{p}^2}{f_\text{z}^2}\delta(\tau^{(m)}) + \pi f_\text{p} \left(1 - \frac{f_\text{p}^2}{f_\text{z}^2}\right) e^{-2\pi f_\text{p} |\tau^{(m)}|}\right].
\end{split}
\label{equation:autocorr_PN}
\end{equation}
From \eqref{equation:autocorr_PN}, we can write the covariance matrix of the phase noise vector $\bm \theta^{(m)}$ as $\left[\bC_{\bm \theta^{(m)} \bm \theta^{(m)}}\right]_{i,j} = R_{\theta^{(m)}\theta^{(m)}}\left( |i-j| \Ts \right)$. From this, it is clear that it the phase noise variance does not depend on the particular time instant at which the phase noise sample is observed, but only depends on the absolute time difference $|i-j|\Ts$. In the following section, our interest lies on estimating the mixed deterministic-random vector of parameters $\bm \xi^{(m)} \triangleq \left[\left\{\bg^{(m)T}[d]\right\}_{d=0}^{D-1},\Delta f^{(m)}, \bm \theta^{(m)}[n], n_0\right]^T$. 

\section{Estimation of synchronization impairments}
In this section, we present a solution to the problem of estimating the parameters in $\bm \xi^{(m)}$. Jointly finding the \ac{ML} estimator for every parameter in $\bm \xi^{(m)}$ is computationally complex. For this reason, we present an approximate solution to the problem of synchronization at low SNR regime. The received signal in \eqref{equation:rx_frame} has \ac{LLF} given by
\begin{equation}
\begin{split}
	&\log p(\{\br^{(m)}[n]\}_{n=0}^{N-1}) \propto -\sum_{n=0}^{N-1}\left\|\br^{(m)}[n]\right\|_2^2 - \\ &- 2\sum_{n=0}^{N-1}\Real\left\{\br^{(m)*}[n] e^{\sfj (2\pi \Delta f^{(m)}n + \theta^{(m)}[n])} \sum_{d=0}^{D-1}\bx^{(m)}[n,d,n_0]\right\} \\ & + \sum_{n=0}^{N-1}\left\|\sum_{d=0}^{D-1}\bx^{(m)}[n,d,n_0]\right\|_2^2.
\end{split}
\label{equation:LLF_time_domain}
\end{equation}

To find the \ac{ML} estimator for $n_0$, we follow the same approach as in \cite{Joint_Sync_Chan_Est_Asilomar_2018}, whereby a low-complexity estimator is given by
\begin{equation}
	\hat{n}_0 = \underset{n_0}{\arg\,\max}\,\sum_{i=1}^{\Lr}\sum_{n=0}^{N-1}\left|r_i^{(m)}[n] s^{(m)\text{C}}[n-n_0]\right|^2.
	\label{equation:delay_mod_corr}
\end{equation}
To compute the modified correlation function in \eqref{equation:delay_mod_corr}, we prepend a $64$-point Golay sequence, which is known to exhibit perfect autocorrelation properties \cite{IEEE:11ad}. This pilot is transmitted with a power $6$ dB larger than the $\Ttr$ OFDM symbols in order to enable frame detection at very low SNR.

Assuming that the timing offset has already been estimated and corrected for in $\br^{(m)}[n]$ to yield $\by^{(m)}[n] = \br^{(m)}[n + n_0]$, we can thereby define
\begin{equation}
\begin{split}
	\phi^{(m)}[n_0,t] &\triangleq e^{\sfj 2\pi \Delta f^{(m)}(n_0 + \Lc + t(K+\Lc))}, \\
	\bE^{(m)} &\triangleq \bigoplus_{n=0}^{K-1}e^{\sfj 2\pi \Delta f^{(m)} n}, \\
	\bP^{(m)}[t] &\triangleq \bigoplus_{n=0}^{K-1}e^{\sfj \theta^{(m)}[n_0 + \Lc + t(K+\Lc) + n]} \\
	\bP^{(m)} &\triangleq \bigoplus_{t=0}^{\Ttr-1}\bP^{(m)}[t] \\
	\bsfS_t^{(m)} &\triangleq \bigoplus_{k=0}^{K-1} \sfs_t^{(m)}[k], \\
	\bsfS^{(m)} &\triangleq \left[\begin{array}{ccc} \bsfS_0^{(m)T} & \ldots & \bsfS_{T}^{(m)T} \\ \end{array}\right]^T \\
	\bsfS_\otimes^{(m)} &\triangleq \bigoplus_{t=0}^{\Ttr-1}\bsfS_t^{(m)} \\
	\bsfg_i^{(m)} &\triangleq [\sfg_i^{(m)}[0], \ldots, \sfg_i^{(m)}[K-1]]^T,\\
	\bv_{t,i}^{(m)} &\triangleq [v_{t,i}^{(m)}[0], \ldots, v_{t,i}^{(m)}[N-1]]^T.
\end{split}
\end{equation}
Then, for the $t$-th OFDM transmitted training symbol and $i$-th RF chain in \eqref{equation:OFDM_t_signal}, $\by^{(m)}[n]$ can be vectorized along the time domain as
\begin{equation}
	\by_{t,i}^{(m)} = \phi^{(m)}[n_0,t] \bP^{(m)}[t] \bE^{(m)} \bF^* \bsfS_t^{(m)} \bsfg_i^{(m)} + \bv_{t,i}^{(m)},
	\label{equation:rx_vect_t_OFDM_i_RF_chain}
\end{equation}
where $\bF$ denotes the $K$-point unitary DFT matrix. Vectorizing \eqref{equation:rx_vect_t_OFDM_i_RF_chain} for the different OFDM training symbols further yields
\begin{equation}
\begin{split}
	\underbrace{\left[\begin{array}{c} 
	\by_{1,i}^{(m)} \\
	\vdots \\
	\by_{\Ttr,i}^{(m)} \\ \end{array}\right]}_{\by_i^{(m)}} &= \underbrace{\left(\bigoplus_{t=0}^{\Ttr} \phi^{(m)}[n_0,t] \bI_{K}\right)}_{\bX[n_0]} \underbrace{\left(\bigoplus_{t=0}^{\Ttr} \bP^{(m)}[t]\right)}_{\bP_\text{E}^{(m)}} \\
	& \left(\bI_{\Ttr} \otimes \bE^{(m)}\right) \left(\bI_{\Ttr} \otimes \bF^*\right) \bsfS^{(m)} \bsfg_i^{(m)} \\ &+ \underbrace{\left[\begin{array}{ccc} \bv_{1,i}^{(m)T} &
	\ldots & \bv_{T,i}^{(m)T} \\ \end{array}\right]^T.}_{\bv_i^{(m)}}
\end{split}
\label{equation:rx_vect_i_RF_chain}
\end{equation}
Now, from \eqref{equation:rx_vect_i_RF_chain}, we can formulate the problem of estimating the parameters in $\bm \xi^{(m)}$ except for the already estimated parameter $n_0$. In the next subsection, we provide the \ac{ML} estimators for the different unknown parameters in \eqref{equation:rx_vect_i_RF_chain}.

\subsection{Joint estimation of CFO, phase noise, and beamformed channels}

In this subsection, we present a joint estimator for the \ac{CFO}, \ac{PN} samples and beamformed channels using the \ac{MAP} criterion. Let $\bm \theta \in \mathbb{R}^{M(n_0 + \Ttr (K + \Lc)) \times 1}$ denote the vector containing the received \ac{PN} samples for the different $\Ttr$ OFDM training symbols and $M$ training frames. It is clear that, to obtain best performance, the complete vector $\bm \theta$ should be estimated from all received measurements corresponding to the different training frames $1 \leq m \leq M$. Such strategy would, however, result in excessive computational complexity. Therefore, we will focus on finding the \ac{PN} vector corresponding to each training frame independently, such that statistical correlation between \ac{PN} vectors for any two different training frames will not be exploited. Under this approximation, the joint negative \ac{LLF} ${\cal L}(\bm \xi^{(m)}) = -\log{p(\by_i^{(m)},\bm \theta^{(m)})}$ of the received vector in \eqref{equation:rx_vect_i_RF_chain} and the \ac{PN} vector for the $m$-th training frame reads as
\begin{equation}
\begin{split}
	{\cal L}&(\bm \xi^{(m)}) \propto \\
	& \frac{1}{\sigma^2}\sum_{i=1}^{\Lr}\left\|\by_i^{(m)} - \bX[n_0] \bP_\text{E}^{(m)} \left(\bI_{\Ttr} \otimes \bE^{(m)} \bF^*\right) \bsfS^{(m)}\bsfg_i^{(m)}\right\|_2^2 \\ &+ \frac{1}{2}\bm \theta^{(m)T}\bC_{\bm \theta^{(m)}\bm \theta^{(m)}}^{-1} \bm \theta^{(m)}.
	\label{equation:MAP_complete_LLF}
\end{split}
\end{equation}
Now, we can obtain the optimum estimator of $\bsfg_i^{(m)}$ by taking the derivative of the objective in \eqref{equation:MAP_complete_LLF} and obtain
\begin{equation}
	\hat{\bsfg}_{i,\text{MAP}}^{(m)} = \frac{1}{\Ttr E_\text{s}}\bsfS^{(m)*} \left(\bI_{\Ttr} \otimes \bF \bE^{(m)*} \right) \bP_\text{E}^{(m)*} \bX^*[n_o] \by_i^{(m)}.
	\label{equation:MAP_beamformed_channel}
\end{equation}
Notice that, owing to absence of prior information, the \ac{MAP} estimator of $\bsfg_i^{(m)}$ coincides with its \ac{ML} estimator. Now, let $\bA^{(m)} \in \mathbb{C}^{\Ttr K \times D}$ be given by
\begin{equation}
	\bA^{(m)} \triangleq \bX[n_0] \bP_\text{E}^{(m)} \left(\bI_{\Ttr} \otimes \bE^{(m)} \bF^*\right) \bsfS^{(m)} \left(\bm 1_{\Ttr} \otimes \bF_1\right),
	\label{equation:equivalent_transfer_matrix}
\end{equation}
with $\bF = [\bF_1, \bF_2]$ being a partition of the DFT matrix, i.e. $\bF_1 \bF_1^* + \bF_2 \bF_2^* = \bI_{K}$. Therefore, we can plug \eqref{equation:MAP_beamformed_channel} into \eqref{equation:MAP_complete_LLF} to obtain the functional
\begin{equation}
\begin{split}
	{\cal L}(\bm \xi^{(m)}) &\propto \frac{1}{\sigma^2 \Ttr E_\text{s}}\sum_{i=1}^{\Lr} \by_i^{(m)} \bA^{(m)} \bA^{(m)*} \by_i^{(m)} \\ &+ \frac{1}{2}\bm \theta^{(m)T}\bC_{\bm \theta^{(m)}\bm \theta^{(m)}}^{-1} \bm \theta^{(m)}.
	\end{split}
	\label{equation:MAP_complete_LLF_PN}
\end{equation}
From \eqref{equation:MAP_complete_LLF_PN}, it is clear that, owing to the non-linear behavior of $\bA^{(m)}$ in \eqref{equation:equivalent_transfer_matrix} with respect to the unknown parameters, optimizing ${\cal L}(\bm \xi^{(m)})$ as a function of $\bm \theta^{(m)}$ in $\bP_\text{E}^{(m)}$ or $\Delta f^{(m)}$ in $\bE^{(m)}$ is a non-convex problem whose solution is very difficult to find, in general. Therefore, we will resort to a suboptimal, yet tractable approximation to solve for $\bm \theta^{(m)}$, and then finally optimize for $\Delta f^{(m)}$. To do this, we can exploit that, mathematically, the \ac{PN} sequence has small amplitude. Therefore, using a first-order Taylor series expansion of $\bp^{(m)} = \vect\{ \diag\{ \bP^{(m)} \}$ is given by $\bp^{(m)} \approx \bm 1 + \sfj \bm \theta^{(m)}$. Then, if we define $\bC^{(m)} \triangleq \bP_\text{E}^{(m)*} \bA^{(m)}$, and $\bY_i^{(m)} = \diag\{ \by_i^{(m)} \}$, we can express \eqref{equation:MAP_complete_LLF_PN} as
\begin{equation}
\begin{split}
	{\cal L}(\bm \xi^{(m)}) &\approx \frac{1}{\sigma^2 \Ttr E_\text{s}} (\bm 1 + \sfj \bm \theta^{(m)})^T \sum_{i=1}^{\Lr} \bY_i^{(m)*} \bC^{(m)} \bC^{(m)*} \bY_i^{(m)} \\ & \times (\bm 1 - \sfj \bm \theta^{(m)}) + \frac{1}{2}\bm \theta^{(m)T}\bC_{\bm \theta^{(m)}\bm \theta^{(m)}}^{-1} \bm \theta^{(m)}.
\end{split}
	\label{equation:MAP_complete_LLF_PN_approx}
\end{equation}

Finally, taking the derivative of the functional in \eqref{equation:MAP_complete_LLF_PN_approx} with respect to $\bm \theta^{(m)}$ yields the optimal $\hat{\bm \theta}_\text{MAP}^{(m)}$ as
\begin{equation}
	\hat{\bm \theta}_\text{MAP}^{(m)} = \left( \Real \{ \bZ^{(m)} \} + 2 \sigma^2 \Ttr E_\text{s} \bC_{\bm \theta^{(m)}\bm\theta^{(m)}}^{-1} \right)^{-1} \Imag\{ \bZ^{(m)} \} \bm 1,
	\label{equation:optimal_PN}
\end{equation}
where $\bZ^{(m)} \in \mathbb{C}^{\Ttr K \times \Ttr K}$ is given by
\begin{equation}
\bZ^{(m)} = \sum_{i=1}^{\Lr} \bY_i^{(m)*} \bC^{(m)} \bC^{(m)*} \bY_i^{(m)}.
\end{equation}
Not surprisingly, the estimator found in \eqref{equation:optimal_PN} is very similar to that of \cite{TSP_PN}. In turn, \eqref{equation:optimal_PN} is the generalization of the optimal \ac{MAP} estimator for $\bm \theta^{(m)}$ for $\Ttr$ OFDM training symbols and $\Lr$ receive RF chains. Finally, the optimal estimator in \eqref{equation:optimal_PN} can be plugged in \eqref{equation:MAP_complete_LLF_PN_approx} to find the optimal \ac{CFO} estimate as
\begin{equation}
\begin{split}
	{\cal L}(\Delta f) \propto \frac{1}{\sigma^2 \Ttr E_\text{s}} \hat{\bp}^{(m)T} \sum_{i=1}^{\Lr} \bY_i^{(m)*} \bC^{(m)} \bC^{(m)*} \bY_i^{(m)} \hat{\bp}^{(m)\text{C}},
	\end{split}
	\label{equation:MAP_complete_LLF_CF}
\end{equation}
where the prior probability density function of the \ac{PN} has been dropped because it does not depend on the \ac{CFO}.

\section{Estimation of high-dimensional frequency-selective mmWave MIMO channel}

In this subsection, similarly to our prior work in \cite{Joint_Sync_Chan_Est_Asilomar_2018} we present an approach to estimate the high-dimensional \ac{mmWave} MIMO channel in the frequency domain. We follow a two-stage estimation strategy in which the \ac{CFO}, \ac{TO}, \ac{PN}, and equivalent beamformed channel are estimated on a frame-by-frame basis. After the transmission of $M$ training frames, these estimates are thereafter used to estimate the MIMO channel. If we define
\begin{equation}
\begin{split}
	\hat{\bsfg}^{(m)}[k] &\triangleq \left[\begin{array}{ccc} 
	\hat{\sfg}_1[k] & \ldots & \hat{\sfg}_{\Lr}[k] \\ \end{array}\right]^T \\
	\bm \Phi &\triangleq \left[\begin{array}{c}
	\bsfq^{(1)T}\bsfFrf^{(1)T} \otimes \bsfD_\text{w}^{(1)-*} \bsfWrf^{(1)*} \\
	\vdots \\
	\bsfq^{(M)T}\bsfFrf^{(M)T} \otimes \bsfD_\text{w}^{(M)-*} \bsfWrf^{(M)*} \\ \end{array}\right],
\end{split}
\end{equation}
we can build the signal model
\begin{equation}
	\underbrace{\left[\begin{array}{c}
	\hat{\bsfg}^{(1)}[k] \\
	\vdots \\
	\hat{\bsfg}^{(M)}[k] \\ \end{array}\right]}_{\hat{\bsfg}[k]} \approx \bm \Phi \vect\{\bsfH[k]\} + \underbrace{\left[\begin{array}{c} 
	\tilde{\bsfv}^{(1)}[k] \\ \vdots \\ \tilde{\bsfv}^{(M)}[k] \\ \end{array}\right]^T}_{\tilde{\bsfv}[k]},
\label{equation:vec_channel_estimates}
\end{equation}
where $\tilde{\bsfv}[k]$ is distributed according to $\tilde{\bsfv}[k] \sim {\cal CN}\left(\bm 0, \left(\bigoplus_{m=1}^{M}\bsfI^{-1}\left(\bsfg^{(m)}[k]\right)\right)\right)$, where $\bsfI\left(\bsfg^{(m)}[k]\right)$ is the \ac{FIM} for the estimation of the vector $\bsfg^{(m)}[k]$. Owing to space limitation, the derivation of the \ac{CRLB} is left for future work. Instead of using the \ac{CRLB} matrix, it is sensible to use the \ac{ML} estimate of the noise variance in \eqref{equation:vec_channel_estimates}, which can be computed using the signal model in \eqref{equation:rx_vect_i_RF_chain} and combining the contributions coming from the different RF chains. The \ac{ML} estimator for the noise variance is computed from the \ac{CRLB} for the estimation of $\hat{\bsfg}[k]$ in \eqref{equation:vec_channel_estimates}. Defining $\bB^{(m)} = \bA^{(m)*}\bA^{(m)}$ allows us to obtain such bound using the \textit{General Linear Model} (GLM) as \cite{Kay:Fundamentals-of-Statistical-Signal:93}
\begin{equation}
	\bsfI^{-1}\left(\{\bsfg^{(m)}[k]\}_{k=0}^{K-1}\right) = \frac{\sigma^2}{K} \bI_{\Lr} \otimes \left(\bsfF \bB^{(m)-1} \bsfF^*\right),
	\label{equation:CRLB_beamformed_channel}
\end{equation}
in which $\Delta f^{(m)}$ can be substituted by $\widehat{\Delta f}^{(m)}$ to obtain the bound.
Then, using the signal model in \eqref{equation:vec_channel_estimates}, we can exploit the extended virtual channel model (see \eqref{equation:channel_extended_fd}) $\vect\{\bsfH[k]\} \approx \left(\tilde{\bsfA}_\text{T}^\text{C} \otimes \tilde{\bsfA}_\text{R}\right)\vect\{\bsfG^\text{v}[k]\}$. Thus, this allows us to use the \ac{SW-OMP} algorithm in \cite{RodGonVenHea:TWC:18}, which has been shown to provide state-of-the-art performance, to estimate the \ac{mmWave} MIMO channel.

\section{Design of hybrid precoders and combiners for joint synchronization and compressive channel estimation}

In this section, we introduce a novel method to design precoders and combiners suitable for joint synchronization and compressive channel estimation at the low SNR regime. The use of several RF chains at the receiver has been shown to enhance estimation performance, especially for the \ac{CFO} parameter \cite{RodGon:CFO_Channel_est:TWC:18}. The main challenge to perform synchronization at low SNR amounts then as to guaranteeing that information coming from $\Lr > 1$ RF chains can be exploited. If the channel response $\bH[d]$ is to be estimated (equivalently, $\bsfH[k]$), we should guarantee that the delay tap $d = d^\star$ at which $\|\bH[d]\|_F^2$ is maximum is preserved after applying a (frequency-flat) precoder $\bsfF$ and combiner $\bsfW$. Mathematically, letting $\bg^{(m)}[d] = \bsfD_\text{w}^{(m)-*}\bsfW_\text{RF}^{(m)*} \bH[d] \bsfF_\text{RF}^{(m)} \bsfq^{(m)} \in \mathbb{C}^{\Lr \times 1}$ be the equivalent channel for an arbitrary training frame $1 \leq m \leq M$, this condition can be written as
\begin{equation}
\underset{d}{\arg\,\max} \|\bg^{(m)}[d]\|_F^2 = \underset{d}{\arg\,\max} \|\bH[d]\|_F^2.
\label{equation:opt_problem_precoders_combiners}
\end{equation}
Optimizing \eqref{equation:opt_problem_precoders_combiners} as a function of the precoders and combiners in $\bg^{(m)}[d]$ is a difficult problem. The resulting precoders and combiners must satisfy the property in \eqref{equation:opt_problem_precoders_combiners} while adopting a sufficiently incoherent design of the resulting measurement matrix $\bm \Phi$ in \eqref{equation:vec_channel_estimates} to estimate the high dimensional MIMO channel $\bsfH[k]$. 

%

In this paper, we propose to combine the proposed Zadoff-Chu training in \cite{ZC-Beam-Training-Myers} with the concept of antenna selection in \cite{RiaRusPreAlkHea:Hybrid-MIMO-Architectures:16}. We will show shortly that, by adopting such design, we enable the use of hybrid architectures without compromising the properties of the measurement matrix $\bm \Phi$ in \eqref{equation:vec_channel_estimates}. Moreover, we shed light on why using Zadoff-Chu sequences with antenna selection is adequate for joint time-frequency synchronization. Let us define ${\cal S}_i = \bigcup_{k=1}^{\Nr/\Lr}(i-1)\Nr/\Lr + k$, ${\cal S}_j = \bigcup_{\ell = 1}^{\Nt/\Lt}(j-1)\Nt/\Lt + \ell$. Let us use $\bH_{i,j}[d] \in \mathbb{C}^{\Nr/\Lr \times \Nt/\Lt}$, $\bH_{i,j}[d] = \left[\bH[d]\right]_{{\cal S}_i,{\cal S}_j}$ to denote the $(i,j)$-th submatrix of $\bH[d]$. Using a partially connected architecture as in Fig. \ref{fig:hybrid_architecture}, with $\Ns = 1$ yet $\Lt \geq 1$, the equivalent beamformed time-domain channel $\bg^{(m)}[d] \in \mathbb{C}^{\Lr \times 1}$ between the hybrid beamformer and combiner may be then expressed as
\begin{equation}
\begin{split}
	\bg^{(m)}[d] &= \left[\begin{array}{ccc}
	\bsfw_1^{(m)*} \bsfH_{1,1}[d] \bsfff_1^{(m)} & \ldots & \bsfw_1^{(m)*}\bsfH_{1,\Lt}[d] \bsfff_{\Lt}^{(m)} \\
	\vdots & \ddots & \vdots \\
	\bsfw_{\Lr}^{(m)*} \bsfH_{\Lr,1}[d] \bsfff_1^{(m)} & \ldots & \bsfw_{\Lr}^{(m)*} \bsfH_{\Lr,\Lt}[d] \bsfff_{\Lt}^{(m)} \\ \end{array}\right] \\
	& \times \left[\begin{array}{ccc} q_1^{(m)} & \ldots & q_{\Lt}^{(m)}\\ \end{array}\right]^T \\
	&= \sum_{\ell=1}^{\Lt} \underbrace{\left[\begin{array}{c} 
	\bsfw_1^{(m)*} \bsfH_{1,1}[d] \bsfff_\ell^{(m)} q_\ell^{(m)} \\
	\vdots \\
	\bsfw_{\Lr}^{(m)*} \bsfH_{1,1}[d] \bsfff_\ell^{(m)} q_\ell^{(m)} \\ \end{array}\right]}_{\bg_\ell^{(m)}[d]},
\end{split}
	\label{equation:Equivalent_Channel_Partial}
\end{equation}
where $\bsfff_j \in \mathbb{C}^{\Nt/\Lt \times 1}$, $\bsfw_i \in \mathbb{C}^{\Nr/\Lr \times 1}$ are the $j$-th and $i$-th hybrid beamformer and combiner employed by the $j$-th transmit and $i$-th receive subarrays. Let $\bsfz_{N}$ denote the $N$-length Zadoff-Chu sequence with root sequence $u$, which is co-prime with $N$. Let us also define $\bsfP_{\text{Tx},\pi_1(m)} \in \mathbb{C}^{\Lt \times \Lt}$ as the permutation matrix obtained by cyclically shifting the $\Lt$ columns of the identity matrix $\bI_{\Lt}$ according to $\pi(m) \in \mathbb{K}$, and $\bsfP_{\text{Rx},\pi_2(m)} \in \mathbb{C}^{\Lr \times \Lr}$ defined similarly.

Since there is no prior information on the MIMO channel $\bH[d]$, it is clear from \eqref{equation:Equivalent_Channel_Partial} that a linear combination of different vectors $\bg_\ell^{(m)}[d]$ leads, in general, to summation of out-of-phase vectors, such that \eqref{equation:opt_problem_precoders_combiners} would not hold. To circumvent this issue, it is clear that turning on a single transmit subarray would accomplish this task. Let us use $j^\star$ to denote the index corresponding to such transmit subarray. In terms of $\bsfff_j$, this means setting $\bsfff_j = \bm 0$, $j \neq j^\star$. Then, similarly to \cite{ZC-Beam-Training-Myers}, we set $\bsfff_j = \bsfP_{\text{Tx},\pi_1(m)} \bsfz_{K_\text{t}}$ for the transmit subarray, with $K_\text{t} = \Nt/\Lt$. Now, \eqref{equation:Equivalent_Channel_Partial} becomes
\begin{equation}
	\bg[d] = \left[\begin{array}{c}
\bsfw_1^* \bH_{1,j^\star}[d] \bsfff_{j^\star} q_j\\
\vdots \\
\bsfw_{\Lr}^* \bH_{\Lr,j^\star}[d] \bsfff_{j^\star} q_j\\ \end{array}\right].
	\label{equation:Equivalent_Channel_Partial_Fixed_Precoder}
\end{equation}
Now, from \eqref{equation:Equivalent_Channel_Partial_Fixed_Precoder}, it only remains to choose the hybrid combiners $\bsfw_i$, $1 \leq i \leq \Lr$. Since performing a linear combination of the different $\bH[d] \bsfff_{j^\star}$ may result in shifting of the maximum of the MIMO channel (in terms of its Frobenius norm), a reasonable strategy is to perform random antenna selection for each receiving subarray. Let us use $p_i^\star \in [1,\Nr/\Lr]$ to denote the index of the antenna selected for the $i$-th subarray. Therefore, we choose $[\bsfw_i]_m = \delta[m - p_i^\star]$, so that the peak of the Frobenius norm of the post-combining channel is preserved.

\section{Numerical Results}

\begin{figure}[!t]
\centering
\includegraphics[width=0.8\columnwidth]{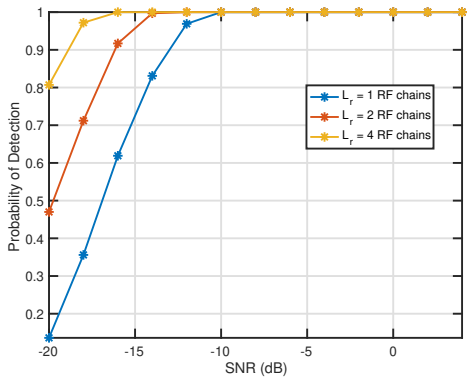}
\caption{Probability of detection as a function of $\SNR$ for $\Lr = \{1,2,4\}$ RF chains.}
\label{fig:Probability_detection}
\end{figure}

In this section, we show numerical results on the proposed \ac{TO}, \ac{CFO} and \ac{PN} synchronization and frequency-selective channel estimation framework. In our simulation setup, the transmitter and receiver are equipped with $\Nt = 128$ and $\Nr = 64$ antennas, and $\Lt = 8$ and $\Lr = 4$ RF chains, respectively. We assume the use of OFDM signaling as in the 3GPP 5G \ac{NR} wireless standard \cite{5G_standard} and \cite{Joint_Sync_Chan_Est_Asilomar_2018}, with $K = 256$ subcarriers and a cyclic prefix length of $\Lc = 64$ to remove ISI for both training and data transmission. We use $M = 32$ training frames, each comprising of $\Ttr = 8$ OFDM training symbols, such that the system overhead is given by $\Ts (\Lc + K)\Ttr M = 42$ $\mu$s. The \ac{CFO} is uniformly generated as $\Delta f^{(m)} \sim {\cal U}(-f_\text{d},f_\text{d})$, where $f_\text{d} = 400$ kHz for illustration. To generate the mmWave frequency-selective channel samples, we use small-scale fading parameters directly obtained from the QuaDRiGa channel simulator \cite{QuaDRiGa_IEEE}, for the 3GPP Urban Microcell (UMi) scenario defined in the 5G channel model \cite{5G_channel_model}, with a Rician factor of $0$ dB. These small-scale fading parameters are thereafter used to generate the MIMO channel according to \eqref{eqn:channel_model}. We show the \ac{NMSE} for the estimation of the \ac{CFO}, and the equivalent beamformed channel. 
We also show the average spectral efficiency obtained using all-digital precoders and combiners to shed light on the best performance that can be achieved using the proposed estimation algorithms. Results are averaged over $100$ MonteCarlo realizations. In Fig. \ref{fig:Probability_detection} the probability of detecting the correct \ac{TO}, which is crucial to perform estimation of the different unknown parameters in $\bm \xi^{(m)}$, as a function of $\SNR$, for $\Lr = \{1,2,4\}$, and $G_\theta = -90$ dBc for the \ac{PN} \ac{PSD}. As we can observe, at the very low SNR regime, the probability of performing time synchronization correctly increases with $\Lr$, while as the SNR increases the value of $\Lr$ tends to be inmaterial since further noise averaging across $\Lr$ received measurements does not further enhance detection performance.

\begin{figure}[!t]
\centering
\includegraphics[width=0.8\columnwidth]{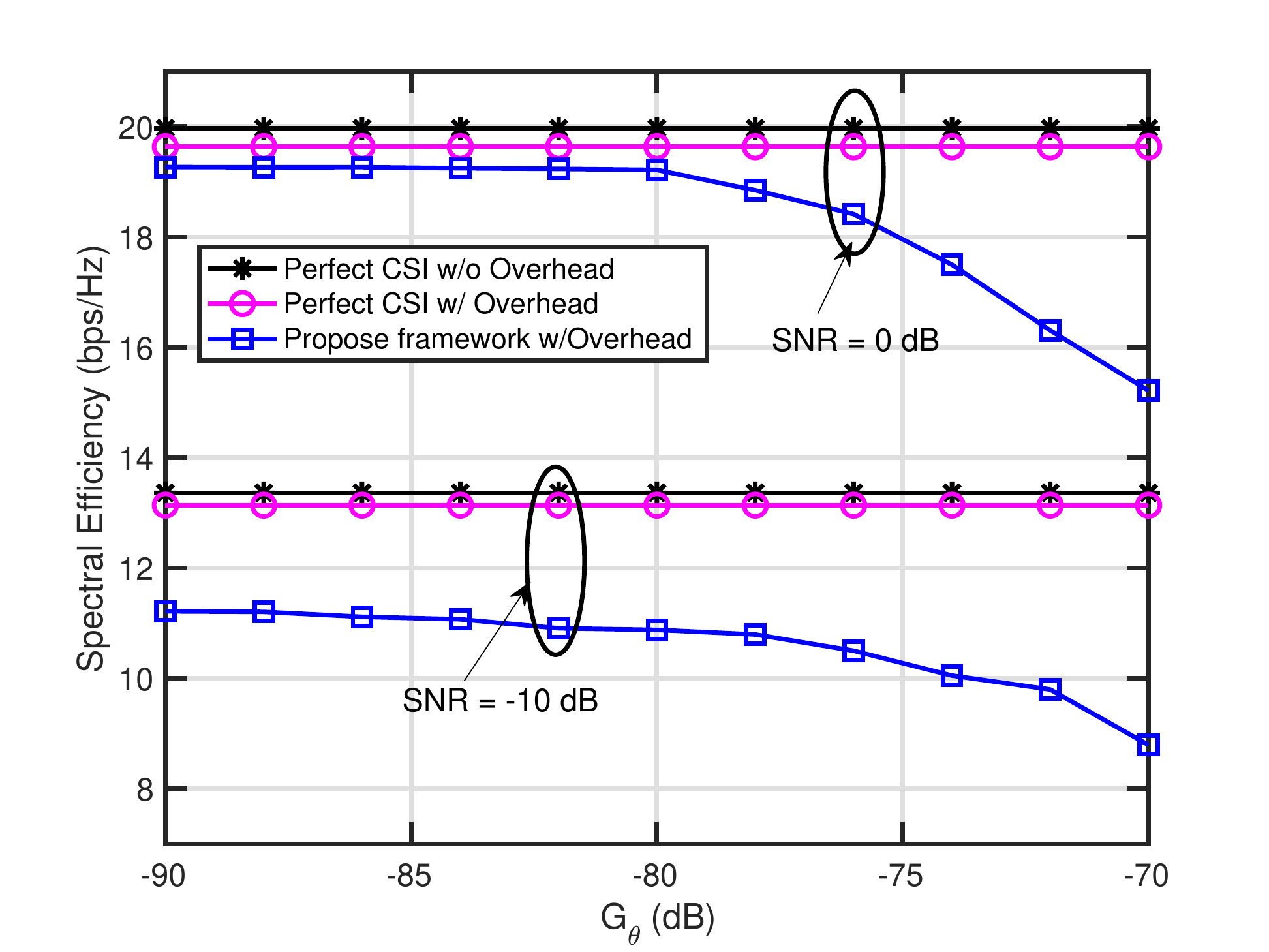} \\
\caption{Evolution of the achievable spectral efficiency as a function of $G_\theta$ in dB units for $\Ns = 2$ transmitted data streams and $\SNR = \{-10,0\}$ dB.}
\label{fig:SE_vs_PN}
\end{figure}

In Fig. \ref{fig:SE_vs_PN}, we show the average spectral efficiency obtained with the proposed estimation framework and the \ac{SW-OMP} algorithm in \cite{RodGonVenHea:TWC:18}, as a function of the \ac{PN} \ac{PSD}, modeled by the parameter $G_\theta$ in \eqref{equation:PSD_PN}, for $\SNR = \{-10,0\}$ dB and for $\Ns = \{1,2\}$ data streams. According to \cite{Blind_PN_ICC_2015_mmWave}, \cite{Draft_IEEE_802.11_PN}, a practical parameter for $G_\theta$ is $G_\theta = -85$ dBc. In our proposed work, we evaluate the performance of the proposed algorithms as a function of $G_\theta$ to gain further insight into the extent to which \ac{PN} sets a bottleneck to the maximum achievable spectral efficiency. For this metric, we design precoders and combiners using the left and right singular vectors of the estimated channels, $\{\hat{\bsfH}[k]\}_{k=0}^{K-1}$, to assess the robustness of the proposed estimation framework and thereby isolate the additional loss incurred owing to hybrid design of the transmit and receive spatial filters.
We also show the upper bound on the spectral efficiency performance taking into account the training overhead, assuming perfect synchronization and \ac{CSI}. To define the total overhead, we measure the dispersion between the MIMO channels experienced by the first and $t$-th transmitted OFDM symbols in terms of \ac{CNMSE} \cite{Joint_Sync_Chan_Est_Asilomar_2018}. For a scenario in which the relative velocity between transmitter and receiver is set to $20$ m/s, and their distance is set to $d = 80$ m, and the channel realizations are obtained using QuaDRiGa channel simulator \cite{QuaDRiGa_IEEE}, a target \ac{CNMSE} of $10^{-3}$ corresponds to a block coherence time of roughly $2.5$ ms \cite{Joint_Sync_Chan_Est_Asilomar_2018}.

\begin{figure}[!t]
\centering
\includegraphics[width=0.8\columnwidth]{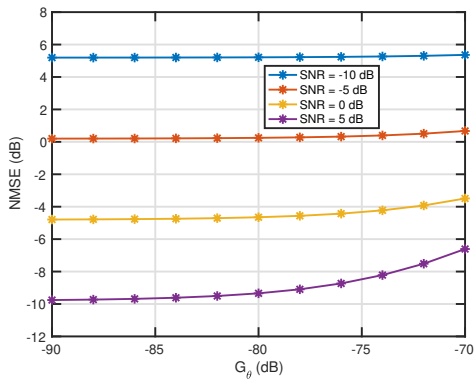}
\caption{Average sample \ac{NMSE} as a function of $G_\theta$ in dB units for $\SNR = \{-10,-5,0,5\}$ dB.}
\label{fig:NMSE_vs_PN}
\end{figure}

If the total number of training samples is $(K + \Lc)\Ttr M \approx 42 \mu$s, the correction factor for spectral efficiency due to training overhead is approximately given by $0.97$. The curves in Fig. \ref{fig:SE_vs_PN} labeled 'w/Overhead' take into account this correction factor. We observe that, as $G_\theta$ increases, the average achievable spectral efficiency decreases, which is the expected behavior. Further, we observe that this behavior is aggravated as the SNR increases, as shown in Fig. \ref{fig:SE_vs_PN}. At very low SNR, the achievable performance is noise-limited, such that a large value of $G_\theta$ does not greatly impact the spectral efficiency up to a certain limit, while at higher SNR the system becomes \ac{PN}-limited, and increasing $G_\theta$ greatly impacts the estimation performance of the equivalent beamformed channels, and the channel matrices themselves, as shown in Fig. \ref{fig:NMSE_vs_PN}.


\section{Conclusions}

In this paper, we developed a joint solution to the problem of \ac{TO}, \ac{CFO}, \ac{PN} and compressive channel estimation for frequency-selective \ac{mmWave} MIMO systems using a frame-wise estimation framework similar to that of 5G \ac{NR}. In spite of the low SNR before configuration of hybrid antenna arrays, synchronization and perfect probability of detecting the transmitted training sequence even at the very low SNR regime. Further, we also showed that, by combining our proposed synchronization framework with the previously proposed \ac{SW-OMP} algorithm, optimum data rates can be attained.

\bibliographystyle{IEEEtran}

\end{document}